# LLM-Feynman: Leveraging Large Language Models for Universal Scientific Formula and Knowledge Discovery

Zhilong Song[1], Qionghua Zhou[1,2,*], Chunjin Ren[2], Chongyi Ling[1], Minggang Ju[1], and Jinlan Wang[1,2,*]

[1]Key Laboratory of Quantum Materials and Devices of Ministry of Education, School of Physics, Southeast University, Nanjing 21189, China

[2] Suzhou Laboratory, Suzhou 215004, China

[*]E-mail: qh.zhou@seu.edu.cn (Q. Z.); jlwang@seu.edu.cn (J. W.)

Distilling underlying principles from data has historically driven scientific breakthroughs. However, conventional data-driven machine learning often produces complex models that lack interpretability and generalization due to insufficient domain expertise. Here, we present LLM-Feynman, a novel framework that leverages large language models (LLMs) alongside systematic optimization to derive concise, interpretable formulas from data and domain knowledge. Our method integrates automated feature engineering, LLM-guided symbolic regression with self-evaluation, and Monte Carlo tree search to enhance formula discovery and clarity. The embedding of domain knowledge simplifies the formula, while self-evaluation based on this knowledge further minimizes prediction errors, surpassing conventional symbolic regression in accuracy and interpretability. Our LLM-Feynman successfully rediscovered over 90% of fundamental physical formulas and demonstrated its efficacy in key materials science applications, including classification of two-dimensional material and perovskite synthesizability and determination of the Green's function and screened Coulomb interaction bandgaps, and prediction of ionic conductivity in lithium solid-state electrolytes. By transcending mere data fitting through the integration of deep domain knowledge, this LLM-Feynman offers a transformative paradigm for the automated discovery of generalizable scientific formulas and theories across disciplines.

## Introduction

The distillation of scientific laws from observation data has historically propelled scientific advancement. One prominent example is Edwin Hubble's groundbreaking measurements of galactic redshifts and distances, which revealed a linear velocity–distance relationship now known as Hubble's law, establishing that the Universe is expanding[1]. More than seven decades later, observations of high-redshift Type Ia supernovae demonstrated that this expansion is accelerating, leading to the discovery of dark energy and fundamentally reshaping modern cosmology[2]. These milestones illustrate how the power of integrating empirical data with domain knowledge through meticulous observation to derive physical formulas and laws with wide-ranging explanatory capability.

Nowadays, we have an explosion of observation data across virtually all scientific domains. Data-driven machine learning (ML) has emerged as an essential tool for identifying underlying patterns[3,4,13,5–12]. While predictive accuracy is a cornerstone of ML, it alone is insufficient for scientific discovery. Interpretability and generalizability are equally critical; models must not only fit the data but also reveal robust, transferable scientific formulas and knowledge. Unfortunately, most ML models operate as black boxes, prioritizing accuracy at the expense of interpretability[14]. In contrast, interpretable ML techniques, such as symbolic regression (SR), systematically explore the landscape of mathematical expressions to identify explicit formulas that minimize prediction errors[14–18]. This approach has proven valuable in material science, successfully uncovering significant scientific insights, including the Gibbs energy of inorganic crystalline solids[19], the tolerance factor for perovskite stability[20], the activity of perovskites in oxygen evolution reactions[8], and the oxide support effects in late-transition metal catalysts[21].

However, a fundamental distinction persists between how machines typically uncover "laws" and how human scientists conduct their work. While data-driven ML can model datasets effectively, it often lacks deep and contextual domain knowledge inherent to human expertise. This domain knowledge is crucial for accelerating the discovery of accurate and interpretable formulas, and extending their applicability across various fields. In contrast, even interpretable ML methods, such as SR, often yield overly complex formula that are challenging to explain and may struggle to generalize beyond their training data[22]. As a result, there is a growing need for methods that effectively integrate domain knowledge with data-driven techniques to derive interpretable and generalizable scientific formulas.

The advent of pre-trained large language models (LLMs) like DeepSeek[23,24], ChatGPT[25] and LLaMA[26] offers unprecedented opportunities to develop tools that seamlessly integrate data with domain expertise, bridging the gap between machine-driven and human scientific reasoning. Trained on vast collections that include scientific literature, textbooks, and databases, LLMs learn comprehensive representations of physics, chemistry, and materials science, thereby transforming various research domains[27]. In materials science, for example, LLMs have facilitated predicting material properties[28], optimizing experimental workflows[29], and proposing synthesis strategies[30–33]. However, leveraging LLMs for the automated discovery of scientific formulas and theories is nontrivial. While these models excel at memorizing and recombining existing knowledge, their autoregressive nature often leads to the generation of plausible-sounding yet physically inconsistent formulas[34,35]. Previous attempts[36] that utilized LLMs for formula discovery, have struggled with effectively incorporating domain knowledge and lacked the multi-objective optimization necessary to balance accuracy, simplicity, and interpretability.

To address these challenges, we developed LLM-Feynman, a framework that combines the scientific reasoning abilities of LLMs with systematic optimization techniques to uncover generalizable, interpretable scientific formula. The framework comprises three primary modules centered around an innovative LLM-based symbolic regression method. It integrates automatic material data preprocessing, feature engineering, and formula interpretation, enhanced by Monte Carlo tree search[37], self-evaluation, and multi-objective optimization. Extensive ablation studies reveal that the integration of domain knowledge and self-evaluation significantly enhances the framework's performance, enabling it to surpass conventional SR methods, such as Sure Independence Screening and Sparsifying Operator (SISSO)[38,39] and PySR[40], in formula accuracy, while maintaining comparable complexity. Notably, LLM-Feynman successfully rediscovered over 90% of the physics formulas from Feynman's lectures. We further applied LLM-Feynman to four critical problems in materials science, including synthesizability of 2D materials and perovskites (classification tasks), and ionic conductivity of lithium solid-state electrolytes and GW bandgap of 2D materials (regression tasks). Our LLM-Feynman paves a way for the automated discovery of generalizable scientific formulas across diverse domains.

## 2. Results and discussion

### 2.1. LLM-Feynman framework

LLM-Feynman consists of three sequential modules (Figure 1): (I) automatic data preprocessing and feature engineering, (II) symbolic regression with self-evaluation and multi-objective optimization, and (III) formula interpretation guided by LLM-based Monte Carlo tree search.

The first module aims to overcome the limitations of raw input data quality and feature relevance, which are critical for the subsequent discovery of interpretable formulas. The input data includes material features $\boldsymbol{X}$, target values $y$ along with the physical meanings and dimensions of both the features and the target. This module automatically preprocesses the data by handling missing values and optionally normalizing features. It employs materials-specific feature engineering through three composition- and structure-driven computational schemes (Figure S1): i) feature selection via mutual information, which retains informative features and removed redundancies using the Automatminer framework[41,42]; ii) LLM-guided feature matching, where the model suggests physically meaningful descriptors retrieved from the Matminer[41] library (template in Figure S2-3); iii) iterative feature refinement, where the feature set is expanded if formula generation stagnates (template in Figure S4). Features derived from these schemes are computed using the Matminer and integrated into the existing feature set. All derived features are optionally enhanced by automatically generated physical meanings and dimensions using the LLM. This customized feature engineering pipeline ensures the discovery of high-quality features, thereby facilitating the generation of accurate and interpretable formulas for various materials science tasks.

The second module addresses the challenge of effectively leveraging the domain knowledge embedded in LLMs while simultaneously optimizing formula accuracy, complexity, and interpretability. It integrates domain knowledge (*i.e.*, physical meanings and dimensions) into the formula generation process through structured prompts (template in Figure S5). Using these prompts, the LLM generates $N$ initial formulas in the form of Python functions. This process is iterative: if fewer than $N$ formulas are generated, LLM continues to produce additional formulas until the desired number is reached. These formulas are required to exhibit high accuracy while also possessing clear physical and chemical interpretability. Accuracy is quantitatively assessed using metrics such as accuracy, MAE and $R^2$. However, interpretability is more complex to quantify. While reduced complexity can improve interpretability, it does not necessarily ensure clear physical or chemical meaning. To address this, we leverage the extensive scientific knowledge of LLM for self-evaluation, assigning an interpretability score S ranging from 0 to 1(template in Figure S6). Subsequently, these formulas are used to calculate their error metrics (MAE, $R^2$ for regression, and accuracy, precision, recall, $F_1$ score, cross-entropy for classification). Combined with complexity C and interpretability score S, a loss function is constructed,

$$L = \alpha N(E) + \beta N(C) + \gamma S$$

Here, N denotes normalization, and the adjustable coefficients $\alpha$, $\beta$, and $\gamma$ (defaulting to 1) represent the equal weighting of accuracy, complexity, and interpretability, respectively. Initially, formulas are scored using the loss function, and formulas (default: 30) with the lowest loss values are selected for further processing. Their mathematical representations, loss values, and associated data (i.e., features, targets, physical meanings, and dimensions) are then incorporated into prompts (template in Figure S7) to iteratively guide the LLM in generating $J$ new formulas (default: 10) in each iteration. These new formulas are self-evaluated and scored using the same loss function. After the specified number of iterations (default: 500), all formulas are consolidated for Pareto frontier analysis, identifying those that strike an optimal balance between accuracy and simplicity.

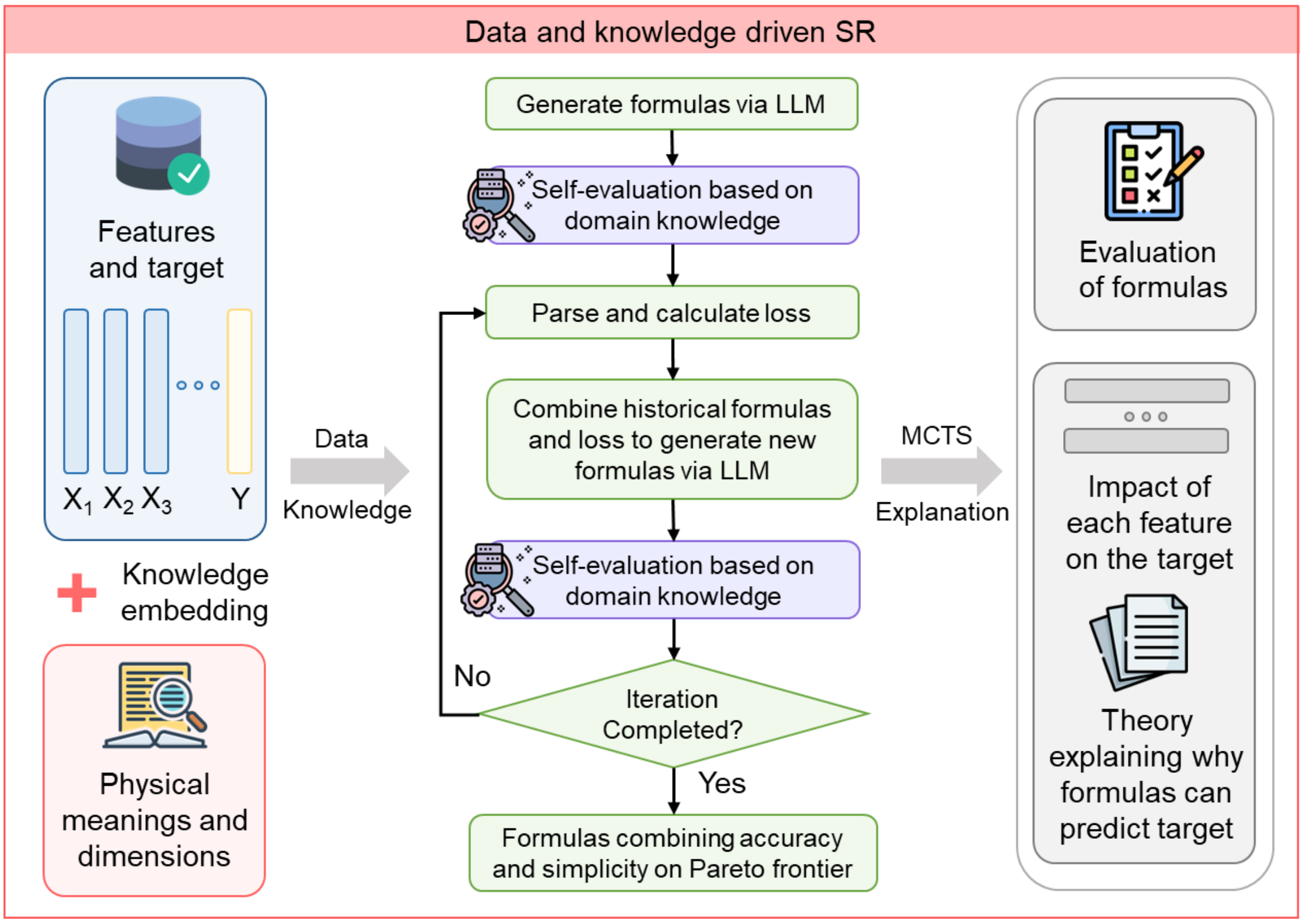

**Figure 1.** Flowchart of the LLM-Feynman framework. (I) The input consists of data, meaning, and dimensionality of features and targets. (II) Formulas are obtained through LLM-based iterative optimization and self-evaluating symbolic regression. (III) Interpretative theories are derived using an LLM-based MCTS framework.

The third module address the critical yet often overlooked challenge of systematically interpreting the physical and chemical meaning of generated formulas. To achieve this, we integrate Monte Carlo Tree Search (MCTS) with an LLM to interpret and refine formula explanations (Figure S7). In this framework, each node in the search tree represents an interpretive hypothesis generated by the LLM, scored by the Upper Confidence Bound (UCB) (see Methods for details). The LLM evaluates the clarity, scientific relevance, and coherence of each hypothesis, updating node scores accordingly. MCTS iteratively explores and refines these explanations, balancing novel ideas with prioritization of high-quality interpretations. This synergy between LLM-driven knowledge and MCTS optimization facilitates efficient navigation of the interpretative space, yielding explanations that are both scientifically meaningful and physically or chemically consistent. Further details are provided in Methods and Supplementary Note S1 and Figure S8-11. For example, Figure S12 illustrates a poor explanation for the formula $\mu/t$ in perovskite OER activity, receiving a self-evaluation score of -72 from the LLM. The explanation fails to meet the core requirement of providing a physical model or theoretical basis for the observed linear correlation between $\mu/t$ and overpotential, merely defining terms instead of offering a complete scientific explanation (Figure S13). In contrast, the explanation refined through MCTS achieves a significantly higher self-evaluation score of 65, effectively resolving these shortcomings (Figure S14).

**2.2. Performance of LLM-Feynman**

To validate the effectiveness of domain knowledge incorporation and self-evaluation, we conducted ablation experiments on the LLM-Feynman framework using density functional theory (DFT) calculation data for single-atom catalysts (SACs). The dataset comprises adsorption energies ($\Delta E$) of intermediates (H, OH, CO, and N) on SACs, with Cu and Ag as support metals and W, Mo, Os, Ru, Ir, Rh, Pt, and Pd as guest elements. Six features were computed for these SACs, including the coordination number of the guest site, valence electron counts of the guest and support metals, electronegativities, and atomic radii. A detailed feature list is available in Table S1.

LLM-Feynman was applied to derive formulas predicting adsorption energies ($\Delta E_{\mathrm{CO}}$, $\Delta E_{\mathrm{OH}}$, $\Delta E_{\mathrm{H}}$, and $\Delta E_{\mathrm{N}}$) based on these six features. We evaluated this under three modes: (I) Data-driven: formulas were discovered without incorporating feature meanings, dimensions, or self-evaluation score (S); (II) Data- and knowledge-driven: built on Mode I, integrating feature meanings and dimensions (domain knowledge); (III) Full LLM-Feynman: combined both domain knowledge and self-evaluation. Three different LLMs (Falcon-Mamba-7B[43], ChemLLM-20B[44], and LLaMA3-8B[45]) were tested, with performance compared to existing data-driven SR methods, SISSO[39] and PySR[40]. For each modality, the average $R^2$, MAE, and formula complexity were assessed for the top 1,000 formulas obtained for each adsorption energy.

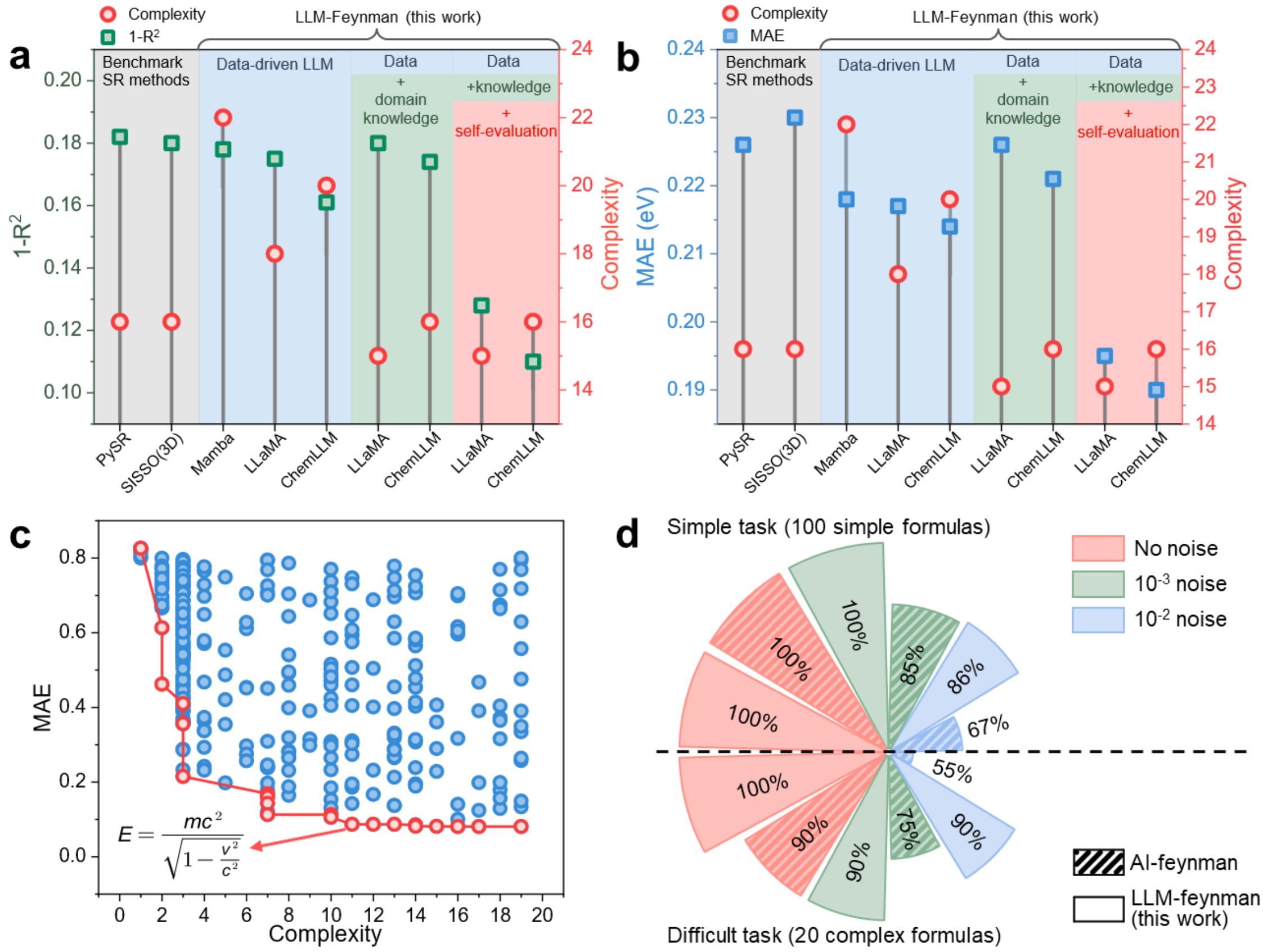

**Figure 2.** Performance of the LLM-Feynman framework. Comparison of **(a)** $1-R^2$ vs. complexity and **(b)** MAE vs. complexity for adsorption energy prediction task among purely data-driven LLM-Feynman, LLM-Feynman with domain knowledge, LLM-Feynman further incorporating self-evaluation, and traditional SR methods SISSO and SR. **(c)** Pareto frontier obtained by LLM-Feynman in deriving the mass-energy equation in special relativity. **(d)** Success rate comparison between LLM-Feynman and AI-Feynman in discovering 100 simple formulas (simple task) and 20 complex formulas (difficult task) from Feynman's Lectures on Physics under different noise levels.

Figure 2a and 2b present a comprehensive ablation study assessing the contributions of each component within our framework. Initially, the data-driven LLM-Feynman framework (devoid of domain knowledge) slightly surpassed baseline methods SISSO and PySR in average $R^2$ and MAE, underscoring the potential of LLMs in SR. However, this advantage was accompanied by increased average formula complexity, suggesting that relying solely on LLM performance may lead to overly complicated expressions. By embedding domain knowledge, LLM-Feynman significantly reduces formula complexity while maintaining competitive accuracy, demonstrating that domain knowledge can steer the model towards simpler, more interpretable formulations aligned to scientific principles. The further addition of self-evaluation, along with domain knowledge, results in notable accuracy improvements over both baseline methods, without increasing formula complexity. This indicates that self-evaluation facilitates iterative refinement, achieving an optimal balance between accuracy, simplicity and interpretability. Among the tested LLMs, Falcon-Mamba-7B showed a substantial performance gap compared to LLaMA3-8B and ChemLLM and was excluded from subsequent tests. Notably, ChemLLM consistently outperforms LLaMA3 across all configurations, likely due to its specialized fine-tuning on chemical and material data, which endows it with comprehensive domain-specific knowledge. This ablation experiment demonstrates the efficacy of embedding domain knowledge and incorporating self-evaluation in leveraging the scientific reasoning capabilities of LLMs, thereby enhancing both accuracy and simplicity of the derived formulas.

In previous ablation experiments, we lacked correct reference formulas for evaluation. To further validate the LLM-Feynman framework's ability to uncover formulas aligned with physical laws, we collected 100 basic formulas from the "Feynman Lectures on Physics" for a straightforward task and 20 complex formulas for a more challenging task. We compared conventional SR method with physical constrains, AI-Feynman[46], with LLM-Feynman utilizing ChemLLM. For both tasks, feature and target data were generated according to their respective formulas, normalized to [0, 1] with Gaussian noise at levels of 0, $10^{-3}$, and $10^{-2}$. The LLM-Feynman framework produces a Pareto frontier, and a formula is correctly identified if it lies on this frontier. Figure 2c presents a Pareto frontier example using LLM-Feynman for deriving the mass-energy formula in special relativity with $10^{-2}$ data noise. The correct formula $E=mc^2$ is successfully identified and located at the lower left corner of the Pareto frontier, where it achieves both high accuracy and low complexity.

Figure 2d illustrates both LLM-Feynman and AI-Feynman successfully identified all formulas in the simple task without noise. However, at $10^{-3}$ and $10^{-2}$ noise levels, LLM-Feynman achieved 100% and 86% success rates, respectively, significantly outperforming AI-Feynman's 85% and 67%. In the more challenging task, the performance gap widened: under $10^{-2}$ noise, LLM-Feynman maintained a 90% success rate, while AI-Feynman dropped to 55%. These results highlight the advantages of LLM-Feynman, largely attributable to the extensive physical and chemical knowledge embedded in pre-trained LLMs. This foundation enables LLM-Feynman to effectively manage complex formulas and noisy data while preserving high accuracy.

### 2.3. Application in the synthesizability prediction of 2D materials and perovskites

After establishing the accuracy, interpretability, and robustness of the LLM-Feynman framework in deriving physically meaningful and simplified formulas, we first applied it to develop formulas for predicting the synthesizability of 2D materials. Synthesizability is critical given the transformative potential of 2D materials in electronics, optoelectronics, energy storage and catalysis. While theoretical studies have predicted approximately 16789 potential 2D materials, supported by databases like Computational 2D Materials Database (C2DB)[47], Materials Cloud[48,49], and 2DMatPedia[50], only around 100 have been experimentally synthesized. This gap underscores the challenge of translating theoretically prediction into experimental reality and highlights the necessity for a formula that can predict synthesizability based on fundamental 2D material characteristics. Currently, no framework specifically addresses this challenge. Two core obstacles impede progress: **(i)** the black-box nature of most machine-learning models, which obscures the underlying physical rationale, and **(ii)** a substantial data imbalance, with very few confirmed synthesizable examples and almost no definitive non-synthesizable cases. These challenges limit interpretability and hinder generalization across chemically diverse systems.

Due to the absence of a curated experimental database for 2D materials, we initially scraped 360 papers using PyPaperBot and employed an LLM to extract reported compositions. Following manual validation and structure reconstruction using ICSD[51] and the aforementioned theoretical databases, we curated 151 synthesized 2D structures as positive examples (Figure 3a). Constructing a corresponding negative set is inherently difficult since unsuccessful syntheses are rarely documented. To solve this, we adopted a Positive–Unlabeled (PU) model using Crystal Graph Convolutional Neural Networks (CGCNN)[52,53]. The accuracy of the PU learning model was boosted from 76.7% to 93.3% by incorporating the Model-Agnostic Meta-Learning (MAML)[54] method (Figure 3b and S16). 150 theoretical candidates with the lowest model output values from the model were classified as non-synthesizable negatives, yielding a balanced dataset (151:150) with a diverse elemental distribution (Figure S15). Additional information on dataset construction and model training is provided in Supplementary Notes S2-3.

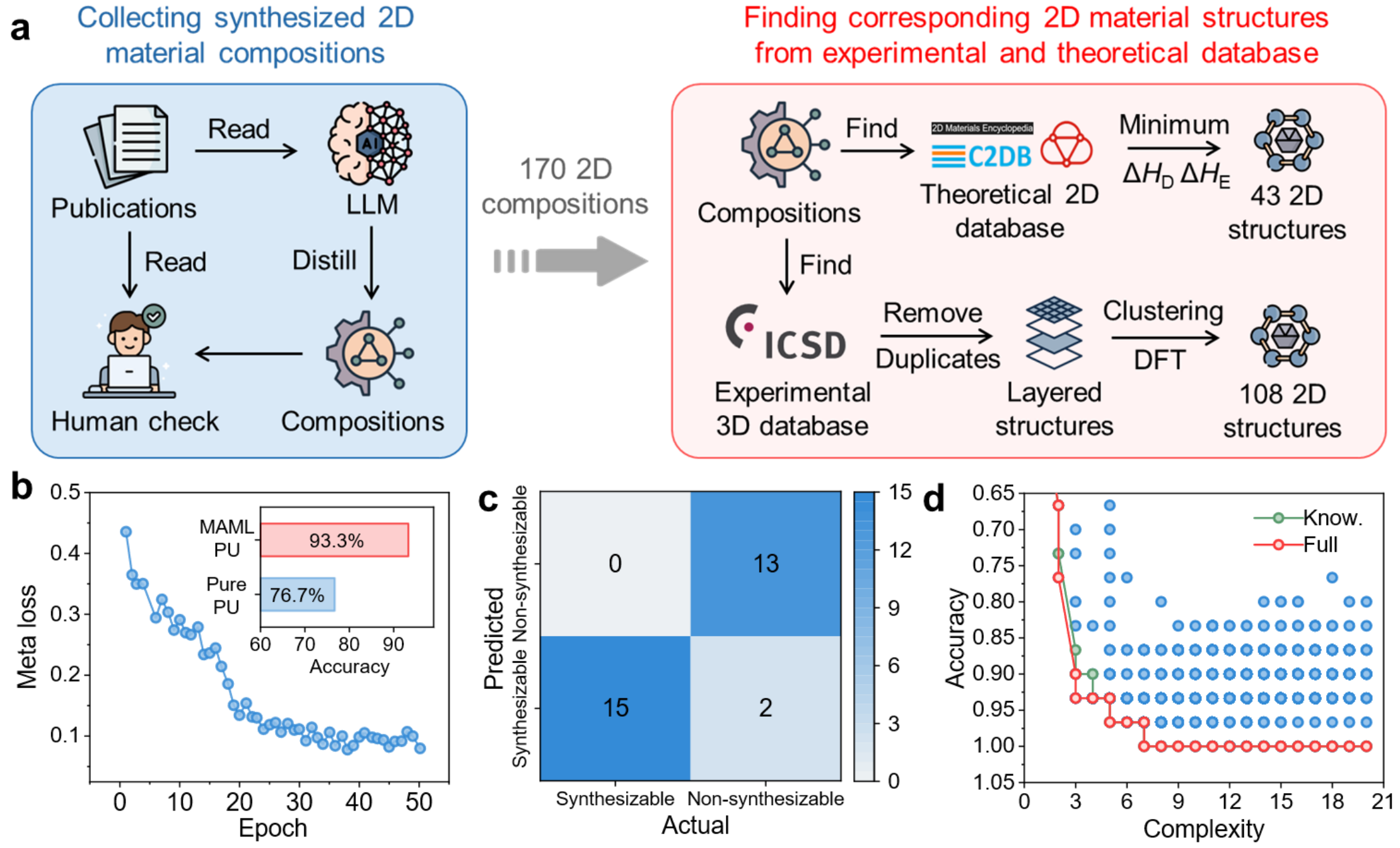

**Figure 3.** Data preparation and performance of LLM-Feynman in the synthesizability of 2D materials. **(a)** Workflow for constructing a database of experimentally synthesized 2D material structures, including LLM-based and manually verified extraction of experimentally synthesized 2D material compositions, as well as structure extraction from experimental and theoretical databases. **(b)** Loss reduction during the training of the PU learning combined with the MAML framework. The inset compares the classification performance of the PU learning model with and without MAML. **(c)** Confusion matrix of the best formula discovered by LLM-Feynman. **(d)** Pareto frontier obtained by LLM-Feynman in exploring the formula for the synthesizability of 2D materials ("Know." and "Full" represent the LLM-Feynman framework that only incorporates domain knowledge and the complete version, respectively).

This dataset was split into training and testing sets with an 8:2 ratio for running the LLM-Feynman framework. Using an automated feature engineering scheme, the Automatminer library initially generated 1,200 structural and compositional features. After applying mutual information screening, 52 features were selected for formula exploration. These formulas underwent Pareto frontier optimization based on four performance metrics (accuracy, precision, recall, and $F_1$ score) as well as formula complexity (Figure S17-19). As depicted in Figure 3d, the green line, representing LLM-Feynman without self-evaluation, shows lower formula accuracy compared to the red line, which includes self-evaluation at equivalent complexity. This highlights the role of self-evaluation in effectively leveraging the scientific knowledge embedded in LLMs, thereby improving formula performance.

From the four Pareto frontiers, two formulas from the lower-left corner were selected for each, resulting in six non-redundant formulas that balance accuracy and complexity (Table S3-4). Among these, the following formula (1), appearing on the Pareto frontier of accuracy vs. complexity (Figure 3d), achieved the highest self-evaluation interpretability scores,

$$\mathrm{MeanUnfilled} - \mathrm{Hal} \times \mathrm{RangeRow} - E_{\mathrm{Hull}} \quad (1)$$

where MeanUnfilled, Hal, RangeRow, and $E_{Hull}$ represent the average number of unfilled *p*-electrons in non-metal elements, the presence of halogen elements, the range of periodic table rows of the elements, and the convex hull energy of formation enthalpy, respectively. This formula misclassified only two synthesizable 2D materials in the test set (Figure 3c). The accuracy, precision, recall, and $F_1$ score reach 0.93, 0.88, 1.00, and 0.94, respectively, significantly surpassing the traditional approach that distinguish synthesizable ($E_{hull}$) and non-synthesizable 2D materials based solely on $E_{hull}$ ($\leq 0.2$ eV/atom and $> 0.2$ eV/atom, respectively; Table 1). This demonstrates that relying solely on thermodynamic $E_{hull}$ for evaluating 2D material synthesizability is insufficient.

**Table 1**. Performance of classification formula for the synthesizability of 2D materials and perovskites on the test set.

| Method | Task | Accuracy | Precision | Recall | $F_1$ score | Complexity |
|---|---|---|---|---|---|---|
| LLM-Feynman | Synthesizability of 2D materials | 0.93 | 0.88 | 1.00 | 0.94 | 7 |
| $E_{hull}$ | Synthesizability of 2D materials | 0.60 | 0.61 | 0.53 | 0.57 | 1 |
| LLM-Feynman | Synthesizability of perovskites | 1.00 | 1.00 | 1.00 | 1.00 | 12 |

The MCTS-guided LLM interpretation reveals that the derived relation effectively captures three key determinants of 2D material synthesizability: electronic structure, chemical composition and thermodynamic stability. A high mean number of unfilled p-electrons favors covalent bonding, thus enhancing structural integrity, while a low formation energy depth ($E_{hull}$) signals an energetically favorable phase, both facilitating successful synthesis. In contrast, the presence of halogens, which complicates reaction pathways and destabilizes intermediates—and a wide range of constituent-element periods, introducing size and electronegativity mismatches, both reduce synthesizability by increasing lattice stress and defect susceptibility. Further details are discussed in Supplementary Note S4. To further validate the framework's applicability to 3D materials, we applied LLM-Feynman to predict perovskite synthesizability, achieving perfect classification performance on the held-out test set (details in Table1, Supplementary Note S4 and S6 and Figure S20).

### 2.4. Applications in the prediction of the conductivity of solid-state electrolyte and GW bandgap of 2D materials

Above results demonstrate that LLM-Feynman excels in both accuracy and interpretability for formula discovery in classification tasks. To further evaluate its general applicability, we applied the framework to two regression tasks: predicting ionic conductivity in solid-state electrolytes and GW quasiparticle bandgaps in 2D materials. Ionic conductivity, which spans over ten orders of magnitude, is highly sensitive to subtle structural motifs and migration barriers, challenging the framework to capture strongly non-linear, thermally activated behavior. In contrast, GW bandgaps are primarily governed by exchange–correlation effects and layer-dependent dielectric screening, representing a complementary challenge in the electronic structure. These tasks explore different length scales, physical mechanisms and data regimes, offering an unbiased evaluation of LLM-Feynman’s capabilities in regression tasks.

To predict the ionic conductivity of solid-state electrolytes, we used an experimental dataset[53] containing 659 data points, divided into training and test sets with a 9: 1 ratio. Using LLM-Feynman, we automatically generated 50 composition features and identified the optimal formula for predicting $\log_{10}\sigma$ via the Pareto frontier analysis (Figure 4a),

$$\sigma = C_1 \cdot (T + {N_{\mathrm{Unfilled}}}^3) + (E_{\mathrm{g}(elem)} - V_{\mathrm{Nd}})^2 - \frac{C_2}{N_{\mathrm{Unfilled}} + E_{\mathrm{g(elem)}}^6} \tag{2}$$

where $T$, $E_{\mathrm{g(elem)}}$, $V_{\mathrm{Nd}}$, and $N_{\mathrm{Unfilled}}$ represent the temperature, average bandgap of constituent elements, average valence electron count of metals, and average unfilled electronic states, respectively. The coefficients $C_1$ and $C_2$ are 0.00403 and 11.8337, respectively. This formula shows good predictive performance with an $R^2$ of 0.855 and an MAE of 0.673 on the testing set (Figure 4b). According to the interpretability module in LLM-Feynman (see Supplementary Note S7 for details), the temperature term reflects thermal activation effects, while the cubic dependence on unfilled states strongly correlates with ion transport. The squared difference between average elemental bandgap and valence electron count captures shifts in conduction pathways. The negative fractional term penalizes excessive unfilled states or large bandgaps that impede carrier mobility. By balancing these factors, the expression accurately predicts ionic conductivity across diverse compositions and temperatures.

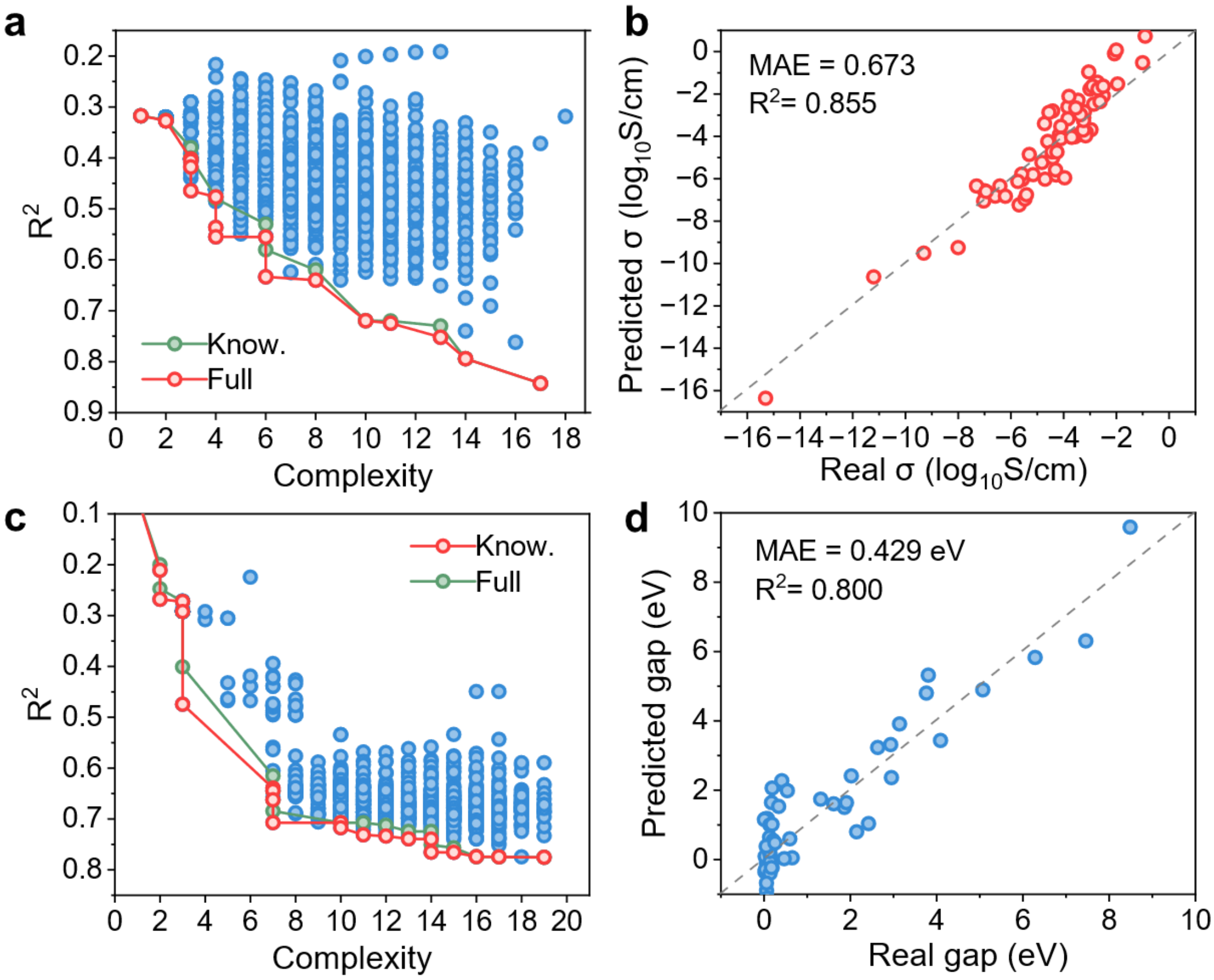

**Figure 4.** Performance of LLM-Feynman on additional classification and regression tasks. **(a)** Pareto frontier and **(b)** predicted vs. actual conductivity for LLM-Feynman in exploring the formula for solid-state electrolyte conductivity. **(c)** Pareto frontier and **(d)** predicted vs. actual GW bandgap for LLM-Feynman in exploring the formula for the GW bandgap of 2D materials.

To predict the GW bandgap of 2D materials, we utilized 551 GW bandgap data from the high-throughput calculations[55] using a 9:1 training-to-test ratio. Our LLM-Feynman identified a Pareto-optimal formula (Figure 4c) from 45 structural component features, successfully capturing the key factors that influence the GW bandgap,

$$E_g = \left(\mathrm{Dist}^3 \cdot (\mathrm{VoroMol} + \mathrm{AtomRadPol} - \mathrm{TetraSys}) - \mathrm{OrthoSys}\right)^6 \tag{3}$$

where Dist represents the structural distance from efficiently-packed atomic arrangements—specifically, how far a material structure deviates from known atomic arrangements that pack together with high efficiency (low packing error). VoroMol denotes the Voronoi coordination number divided by molecular volume, and AtomRadPol is the atomic radius divided by polarizability. TetraSys and OrthoSys are binary indicators for tetragonal/orthorhombic crystal systems, respectively (1 if true, 0 otherwise). This formula achieves an $R^2$ of 0.80 and an MAE of 0.429 eV on the testing set (Figure 4d and Supplementary Note S8). The predictive power of this formula stems from its four-step hierarchical design: First, by cubing the distance term ($Dist^3$), the model heavily penalizes structures that deviate significantly from well-packed structures, ensuring that materials with poorly arranged atoms contribute minimally to the prediction. Second, it incorporates two fundamental descriptors: atomic packing density (VoroMol) and electronic polarizability (AtomRadPol), which capture both geometric constraints and electronic flexibility that govern band gap behavior. Third, the formula accounts for crystal symmetry effects through systematic adjustments: a moderate correction for tetragonal systems and a larger adjustment for orthorhombic lattices, reflecting their characteristic electronic structures. Finally, raising the entire expression to the sixth power creates an exponential sensitivity that amplifies even subtle structural or symmetry differences into measurable changes in the predicted band gap.

**Table 2**. Performance comparison of three feature computation methods in LLM-Feynman on two tasks.

| Feature computation methods | Solid-State Electrolyte Conductivity Test Set $R^2$ | 2D Material GW Bandgap Test Set $R^2$ |
|---|---|---|
| LLM-suggest | 0.687 | 0.711 |
| LLM-suggest-iterative | 0.709 | 0.715 |
| Automatminer+MI | 0.855 | 0.800 |

Having confirmed that LLM-Feynman produces accurate and interpretable formulas for both regression tasks, we next tested whether the feature-generation step could also be automatically provided by the LLM. In addition to the default Automatminer feature engineering pipeline, which selects features by mutual-information ranking, we tested two LLM-based approaches (Table 2): i) LLM-suggest, where a single prompt asks the model to propose an initial feature set, and ii) LLM-suggest-iterative, where the LLM provides new features when performance of formulas derived by LLM-Feynman shows no improvement for 50 iterations. Across both ionic-conductivity and GW-bandgap datasets, formulas built from Automatminer features performed better; those based on the LLM suggestion show the weaker performance, while the LLM-suggest-iterative approach only partially bridges the gap. Therefore, while the current LLMs can effectively integrate high-quality features, they are not yet competitive with existing automated tools in the task of creating features from scratch. This indicates that improvements are needed either in LLM-guided feature engineering or in the LLMs themselves.

## Conclusion

In summary, we have developed the LLM-Feynman framework, an innovative methodology that seamlessly integrates automated data preprocessing, LLM-based symbolic regression with self-evaluation and multi-objective optimization, and Monte Carlo tree search for formula interpretation. This framework automatically extracts generalizable and interpretable scientific formulas from data and domain knowledge. It outperforms conventional methods like SISSO and PySR in accuracy and complexity, robustly rediscovering over 90% of the 120 physical formulas from Feynman's physics lectures. The framework demonstrates exceptional performance, achieving over 90% accuracy in predicting the synthesizability of 2D materials and perovskites and achieving with $R^2$ values above 0.8 in estimating ionic conductivity and GW bandgaps. Its modular design, compatible with any state-of-the-art LLM, positions it as a transformative tool that will evolve with advancements in LLM technology, driving future breakthroughs in materials science. This LLM-Feynman framework bridges the gap between domain knowledge and data-driven, interpretable ML methods, paving the way for fully automated, high-fidelity scientific discovery.

Despite these advances, our framework is currently limited by the token-length restrictions of contemporary LLMs, which impede its application to tasks involving large datasets (e.g., on the order of $10^4$ samples). Addressing this bottleneck is essential for extending its applicability to large-scale problems. Future efforts will focus on solutions such as hierarchical data processing, architectural enhancements, and the incorporation of reinforcement learning with human feedback[56] to effectively manage and optimize large inputs.

# 3. Methods

## 3.1 Inferences of large language models

Three large language models, Falcon-Mamba-7B[43], ChemLLM-20B[44], and LLaMA3-8B[45] were employed in the LLM-Feynman framework. These models were loaded using the Transformers library, with inference accelerated by FlashAttention[57] and Accelerate library[58]. We adjusted key decoding parameters, setting the temperature to 0.7, top-k to 50, and top-p to 0.9 to balance diversity and precision. Figures S2-7,9-14 shows the model prompts, designing to provide structured responses in LLM-Feynman. All inferences were performed on a server with one NVIDIA A800 GPU with 80 GB VRAM and 1 TB memory. The code of LLM-Feynman is shown in https://github.com/szl666/LLM-formula.

## 3.2 Feature engineering

To automate feature engineering for materials science formula discovery, we employed Automatminer[41,42], a framework that integrates multiple feature generation strategies to construct an initial feature set, denoted as $\boldsymbol{X}_{\text{ini}}$. Automatminer systematically extracts formulas from composition, structure, and electronic properties, ensuring a comprehensive representation of material characteristics. Feature selection was achieved via mutual information (MI), implemented using the MODNet[59] preprocessing module. MI quantifies the dependence between two variables, defined as,

$$I(X,Y) = \sum_{x \in X} \sum_{y \in Y} p(x,y) \log \frac{p(x,y)}{p(x)p(y)}$$

where $p(x, y)$ is the joint probability distribution, and $p(x)$, $p(y)$ are marginal distributions. MI is computed between features and the target variable to retain informative formula, while redundant features with high MI among themselves are removed. This results in the final optimized feature set, $\boldsymbol{X}_{\text{ifinal}}$, enhancing model interpretability and predictive performance.

### 3.3 MCTS explanation

In MCTS explanation based on LLM module, each tree node represents an interpretive idea proposed by the LLM, with its score determined by the Upper Confidence Bound (UCB),

$$UCB = Q + c\sqrt{\frac{\ln N_{\text{parent}}}{N_{\text{node}}}}$$

where $Q$ is the self-evaluation score, $N_{\text{parent}}$ and $N_{\text{node}}$ denote the visit counts of the parent and current node, respectively, and $c$ balances exploration and exploitation.

The module operates in four iterative stages: Selection, where nodes with the highest UCB values are chosen; Expansion, where the LLM generates refined interpretations; Self-evaluation, where the LLM scores new nodes for clarity and relevance; and Backpropagation, where scores update ancestor nodes. This synergy between MCTS and LLM enables efficient exploration of scientifically meaningful interpretations.

## Acknowledgements

This work was supported by the National Key Research and Development Program of China (grant 2021YFA1500700), the Natural Science Foundation of China (grant 22033002, 22373013, T2321002), and the Basic Research Program of Jiangsu Province (BK20232012, BK20222007), Jiangsu Provincial Scientific Research Center of Applied Mathematics (BK20233002) and the Fundamental Research Funds for the Central Universities. We thank the Big Data Computing Center of Southeast University for providing the facility support on the calculations.

## Conflict of Interest

The authors declare no competing interests.